# Examining the Cognitive Gap Between Authors and Peer Reviewers on Academic Paper Novelty

Chenggang Yang and Chengzhi Zhang*

Department of Information Management, Nanjing University of Science and Technology, Nanjing, 210094 China
ichigo@njust.edu.cn, zhangcz@njust.edu.cn

**Abstract.** Novelty is a crucial metric for assessing the quality of academic papers. Scholars strive to highlight the novel aspects of their work, particularly in the title, abstract, and introduction. Peer review, serving as the gatekeeper of scientific rigor, rigorously evaluates the novelty of papers, yet a cognitive gap may exist between author self-promotion and reviewer evaluation. To investigate this, we analyzed 15,328 academic papers published in Nature Communications from 2016 to 2021, along with their peer-review comments. We found that both reviewers and authors emphasize result-oriented innovation, with reviewers adopting a more comprehensive evaluation perspective. Furthermore, by examining promotional intensity against inherent paper novelty, we discovered its effect is depend on the paper's actual innovation level. Highly innovative papers benefit from stronger promotional language, receiving more positive evaluations. We also found that promotional language significantly correlates with reviewer disagreement on novelty specifically for papers of moderate innovativeness, whereas it has negligible impact for papers with either very high or very low novelty. This reveals how promotional language operates most prominently in the gray area of academic evaluation.

## 1 Introduction

In recent years, the number of academic papers has grown exponentially. To ensure that high-quality research is published promptly and accurately in appropriate journals or conferences, the pressure on peer reviewers, who serve as gatekeepers of scientific publishing, has intensified. However, peer review is not without its challenges, including inefficiencies and potential biases(Parker et al., 2018; Stelmakh et al., 2019; Wicherts, 2016). Peer review typically evaluates manuscripts along multiple dimensions, including novelty, methodological rigor, and clarity of presentation. Among these criteria, novelty plays a particularly prominent role in many editorial and reviewer assessments, especially in high-impact journals.

* Corresponding author.

The ability to communicate novel ideas and research findings effectively is an indispensable part of academic research and is crucial across many scientific domains, such as grant applications, patent writing, and academic paper writing (Peng et al., 2024). Academic papers are the primary medium for disseminating research outcomes, enabling researchers to share their discoveries and insights. Clear articulation of a study's contribution can facilitate the efficient communication of novelty, even among expert readers. To accurately and efficiently convey the innovative aspects of their work, researchers often highlight their contributions in the title, abstract, and introduction of their papers. These promotional statements have been shown to correlate with the subsequent impact of the papers(Pearson, 2020; Wheeler et al., 2021).

However, inappropriate promotion in academic papers can lead to adverse consequences. Some scholars, due to insufficient consideration of prior literature, may exaggerate the novelty of their work. If the research findings are later found to be less novel than claimed, the authors' academic reputation may be affected, and other researchers may be misled into investing time and resources in less productive directions. Conversely, some scholars may insufficiently promote or inaccurately describe the novelty of their research, which may cause reviewers to overlook or underestimate the innovative aspects of the work. In such cases, reviewers may question exaggerated claims or request revisions when they detect inconsistencies between the stated contributions and existing literature. As shown in Table 2 in Appendix, if exaggeration is detected, reviewers may point out the exaggerations in their comments, such as "*Smith et al. (2021) demonstrated that while advancements in data privacy are significant, they often come with trade-offs in model performance and complexity.*", and provide corresponding references as evidence for the authors to revise their papers. Similarly, insufficient promotion may make it more difficult for reviewers to clearly identify the novel contributions of the research, which may influence their overall evaluation of the manuscript. For papers with insufficient promotion, reviewers may struggle to grasp the innovative aspects of the research, leading to the findings being undervalued. This can result in a lower overall evaluation of the paper's quality by reviewers, ultimately affecting its chances of publication.

Therefore, appropriate promotion is crucial for reviewers to make informed decisions regarding acceptance or rejection. Specifically, we address the following five questions:

Drawing on Leahey et al.(2023), we distinguish three types of innovation in academic work: theoretical, methodological, and results innovation. The definitions and descriptions of these three types of novelty are presented in Table 3. We apply this framework to both sides of the review exchange, classifying authors' self-promotion and reviewers' assessments into theoretical, methodological, and results-oriented novelty evaluations. This motivates our first question:

***RQ1:*** Which aspects of innovation do paper authors and reviewers prioritize more?

Addressing RQ1 allows us to quantify where authors and reviewers place emphasis when presenting or apprasing novelty. Building on these measures, we next examine alignment versus divergence between the two sides:

***RQ2***: What are the differences in focus between paper authors and peer reviewers regarding the innovation of a paper?

After establishing the existence and nature of the cognitive gap, we delve deeper into the factors that influence authors' promotional behavior. We hypothesize that the use of promotional language is not arbitrary but is strategically coupled with the paper's inherent novelty, leading us to propose RQ3:

***RQ3***: What is the relationship of inherent novelty of a paper and the author's use of promotional language?

Subsequently, we examine the consequences of this promotional behavior. Specifically, we investigate whether the intensity of promotional language uniformly influences reviewers' perceptions or if its effect is depended on the paper's inherent qualities. So, we propose RQ4:

***RQ4***: How does the intensity of promotional language influence the reviewers' evaluation of novelty?

Finally, we investigate process-level variation. Reviewers may differ in how they interpret and weigh promotional cues. We therefore examine whether promotional intensity amplifies or attenuates disagreement among reviewers:

***RQ5***: What is the relationship between the intensity of promotional language and the disagreement of reviewers in novelty assessment?

The primary contributions of this paper are manifested in the following three aspects:

Firstly, we provide systematic quantification of the aspects of innovation prioritized by paper authors versus peer reviewers. It precisely characterizes the alignments and divergences in their focus.

Secondly, we investigate not only how inherent novelty drives authors' use of promotional language but also establishes a complete pathway from a paper's intrinsic qualities to author behavior to reviewer perception.

Finally, we move examines the impact of promotional language on the stability of the review process itself. It investigates whether such language amplifies or attenuates disagreement among reviewers.

## 2 Related work

The related work section of this study encompasses two parts: research on promotional language in academic papers and research on innovation evaluation in peer reviews.

### 2.1 Promotional language

Promotional language refers to the linguistic expressions and stylistic choices designed to market or advocate research findings. This type of language is often characterized by exaggeration, subjectivity, or emotional appeal, which may influence the reader's objective understanding of the research. Previous studies on promotional sentences have primarily concentrated on grant or project proposals. For instance, Millar et al.(2022) analyzed 717 NIH grant applications and found that applicants increasingly describe their work subjectively, relying on promotional language and emotional appeals. Peng et al. (2024) examined the promotional language in funding applications from NIH, NSF, and the Nord Foundation, investigating its relationship with the likelihood of funding and the future impact of the projects. It is evident that the dissemination of research is also closely related to the manner in which it is promoted.

**Table 1.** Related Work of Promotion Language in Academic article

| Source | Authors | Main findings |
|---|---|---|
| Title | Pearson(2020) | The study found that the structure and characteristics of titles may influence a paper's academic impact. |
| | Sagi & Yechiam(2008) | The study provides empirical evidence that humorous titles in scientific articles are associated with fewer citations. |
| | Jiang & Jiang, (2023) | Reveal significant trends in title length, complexity, and syntactic structures. |
| Abstract | Wheeler et al.(2021) | Reveal a significant increase in the use of personal pronouns and expressive confidence (referred to as "clout") in psychology journal abstracts. |
| | Li(2022) | It explores the relationship between passive voice usage and active voice initiated by personal pronouns, contributing to a better understanding of the evolving style of academic writing. |
| | Song et al.(2023) | The findings suggest that papers published in higher quartile journals tend to exhibit greater lexical density and sophistication, implying a connection between writing quality and scientific impact. |

Current research on promotional language in academic papers delves into the titles, abstracts, and main text content, aiming to uncover phenomena present in scientific research and to provide recommendations for academic writing, thereby enhancing the quality of scholarly articles. Citation counts are often used as a proxy for the influence of a paper to study the effects of promotional language. Titles and abstracts, serving as summaries of the entire paper, are common corpora for research on promotional language in papers. Metrics such as length, vocabulary usage, and semantic complexity have been extensively studied by many researchers(Jiang & Jiang, 2023; Li, 2022; Pearson, 2020; Sagi & Yechiam, 2008; Wheeler et al., 2021), as detailed in Table 1. However, the main text of academic papers remains under-researched due to the challenges in data acquisition and processing. In recent years, the open access to large-scale paper datasets and the development of large language models have propelled research in the semantic understanding of full-text academic papers. Such research often correlates writing style with other metrics. For example, Lu et al.(2019) found that cultural background influences sentence structure and word choice. Costello et al. (2023) examined the use of uncertain language (e.g., “may,” “might,” or “possibly”) in papers written by male and female authors and found a relationship between gender and writing style, as well as evidence that editors play an important role in mitigating these differences. Wu et al.(2025) found that the introduction part of the paper is more suitable for measuring its novelty.

Moreover, there is relatively scant research utilizing large-scale corpora to investigate the use of promotional language in academic papers. Thanks to the advancements in large language models, we are now better equipped to process the vast corpora that are currently available, extracting more information from them. This study draws on research related to promotional language in grant applications, employing large language models to automatically extract and classify promotional language from the introductions of academic papers. Based on this, we assess the intensity of promotional language in academic papers and examine the relationship between the intensity of promotion and the level of endorsement by reviewers. Compared with studies on grant proposals, relatively little research has examined how authors highlight and communicate novelty in published scientific articles and how such promotion relates to peer-review evaluations.

### 2.2 Innovation Evaluation in peer review

Peer review stands as the cornerstone of academic exchange and the bedrock of scientific publishing(Ghosal et al., 2022). Peer experts, with their profound domain knowledge and professional experience, should be capable of evaluating the overall quality of academic papers. The innovativeness of a paper, being a decisive factor of its quality, has always been highly regarded by reviewers(Teplitskiy et al., 2022). However, the peer review process has been subject to controversy due to its lengthy review cycles, lack of transparency, and potential biases (Parker et al., 2018; Stelmakh et al., 2019; Wicherts, 2016). To address these issues, several journals, including Nature Communications and Plos One, have begun to make peer-review comments publicly available ("Transparent Peer Review for All," 2022). The disclosure of peer-review comments has promoted transparency in the review process and enhanced the efficiency of communication between paper authors and reviewers. Reviewers are expected to make rational judgments about the quality of papers and provide revision suggestions to authors based on these judgments, without being influenced by other factors. For instance, Bornmann et al.(2019) performed a crucial validation study, comparing bibliometric novelty indicators against expert assessments provided by F1000Prime. Wu et al. (2025) fully leveraged peer-review comments and proposed an automated approach for evaluating the novelty of academic paper methods, which integrates human expertise with large language models.

To enhance researchers' understanding of innovation evaluation during the review process, we employed rule-based and machine learning methods to extract innovation evaluation sentences from peer-review comments. We studied the current methods and focus points of innovation evaluation in the review processes across different disciplines and, in conjunction with the promotional language in the introductions of papers, provided recommendations for authors on the use of promotional sentences in academic writing.

## 3 Data and Methodology

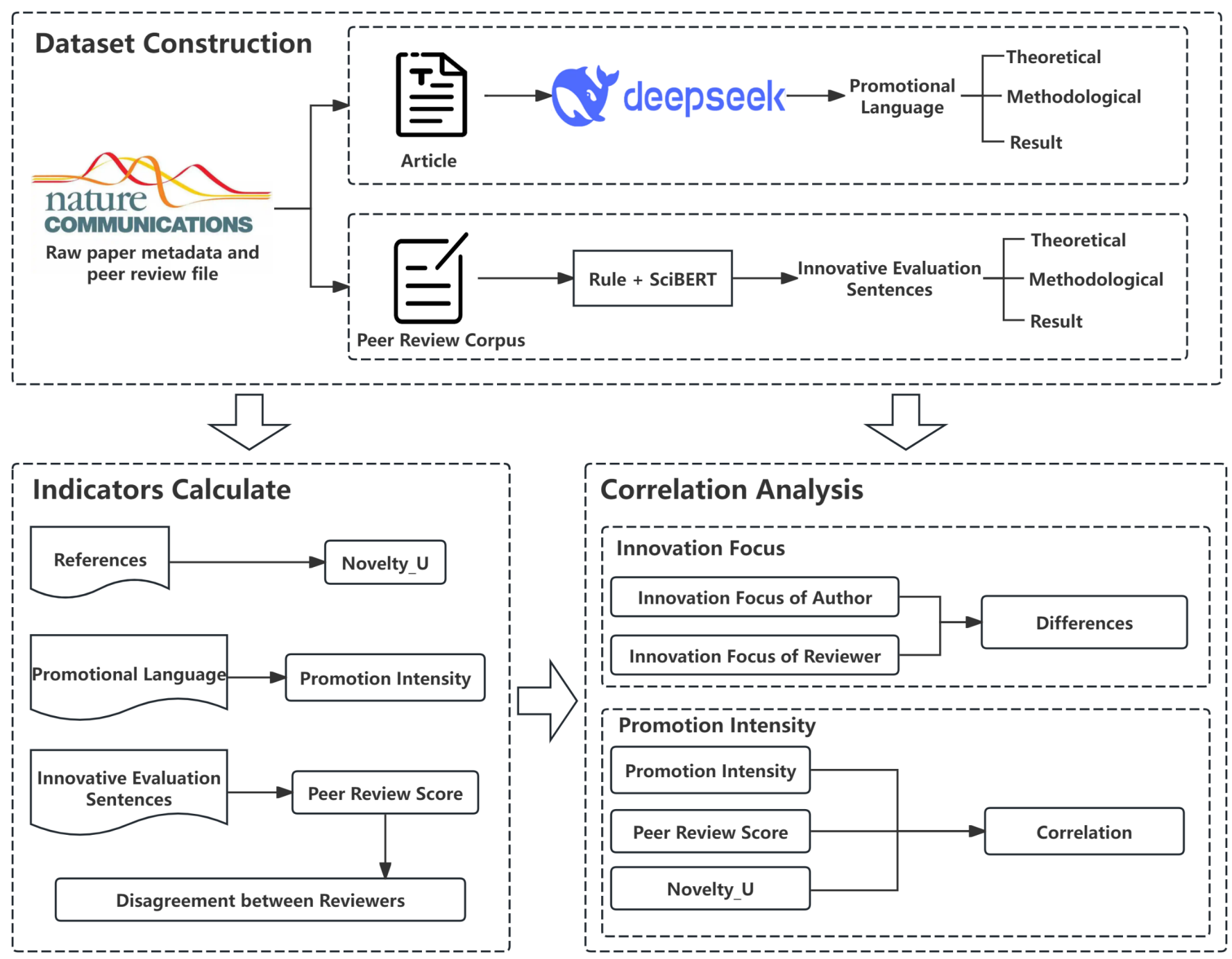

**Figure. 1.** Framework of this Study

The aim of our study is to examine the differences in focus between paper authors and reviewers regarding the novelty of papers during the publication process, as well as the relationship between the use of promotional language by authors and the evaluations by reviewers. To achieve this, we utilized original academic papers and peer-review comments from various fields in *Nature Communications*, extracting sentences related to innovation evaluation for analysis. We also assessed the promotional intensity of the original papers using the novelty metric proposed by Uzzi et al.(2013). The framework of our study is illustrated in Figure 1. Specifically, we conducted our research in three steps. The first step involved the construction of the dataset, where we collected all academic papers published in *Nature Communications* from 2016 to 2021 along with their publicly available peer-review comments. We parsed their contents to extract authors' promotional language about their own research from the introduction of original papers and reviewers' innovation evaluation sentences from the review comments. The second step was the comparison of innovation focus points across different disciplinary fields. Based on the five disciplinary categories provided by the *Nature Communications* website, we observed how researchers' focus on innovation varies across fields and how the focus points of paper authors and reviewers differ regarding the novelty of papers. The third step combined a reference-based method for calculating paper novelty to investigate the relationship between the use of promotional language in papers with different levels of novelty and reviewer comments.

### 3.1 Dataset and Data Preprocessing

This section outlines the process of dataset construction and preprocessing for our study. Initially, we collected the original papers and peer review comment files from the *Nature Communications*[1] website. Subsequently, we employed large language models to extract promotional language from the original papers and utilized a "rule-based + machine learning" approach to extract innovation evaluation sentences from the peer-review comments. This groundwork lays the foundation for our subsequent analysis of the authors' and reviewers' evaluations of the papers' innovativeness.

**Raw article and peer review corpus collection**. The data source for this study is Nature Communications, a subsidiary journal of Nature. This journal encompasses the latest research findings across various fields of natural sciences and has been committed to the transparency of peer review to enhance the quality of the review process, being one of the earliest journals to make peer-review comments publicly available. Since 2016, authors have had the option to disclose the exchanges between themselves and the reviewers. Papers with disclosed review comments can be found with corresponding peer review PDF files on their content pages, which include the reviewers' comments and the authors' responses from each round of review.

**Table 2.** Disciplinary Distribution of Original Dataset

| Disciplinary \ Year | 2016 | 2017 | 2018 | 2019 | 2020 | 2021 |
|---|---|---|---|---|---|---|
| Biological science | 19 | 984 | 2142 | 2241 | 2559 | 794 |
| Earth and environmental sciences | 10 | 64 | 149 | 55 | 221 | 55 |
| Health sciences | 30 | 155 | 395 | 460 | 577 | 219 |
| Physical sciences | 12 | 431 | 1003 | 828 | 1187 | 376 |
| Scientific community and society | 5 | 5 | 30 | 36 | 47 | 17 |

We collected the publication dates, titles, abstracts, main texts, and publicly available peer review PDF files of all papers published from 2016 to 2021 from the *Nature Communications* website. We collected the publication dates, titles, abstracts, main texts, and publicly available peer review PDF files of all papers published from 2016 to 2021 from the *Nature Communications* website. Since manuscripts may undergo multiple rounds of peer review, the novelty evaluation score in this study is calculated based on reviewers' comments from all available review rounds. To examine whether the novelty of a paper changes substantially during revision, we randomly selected 50 papers from our dataset and retrieved their original preprint versions from *arXiv* or *bioRxiv* when available. By manually comparing the titles, abstracts, and reference lists, we found that in more than 80% of the cases the novelty-related content remained largely unchanged between the preprint and the final published versions. Based on the journal's disciplinary classification, we categorized the papers into five fields: biological science, health science, earth and environmental science, physical science, and scientific community and society. A single paper could belong to multiple disciplinary fields. We identified the structure of the papers using HTML tags within the main text and extracted the introduction sections as the corpus for subsequent promotional language extraction. We required that the collected papers have complete titles, authors, abstracts, and introduction content, along with publicly available peer-review comments. Ultimately, we gathered 15,328 academic papers along with their peer-review comments. The distribution of the number of papers in each discipline is shown in Table 2. Since the reviewers' comments and authors' responses in the publicly available peer-review comments for each paper were contained within the same PDF file, we used the Python package PyMuPDF[2] to parse the text and segmented the reviewers' comments from the authors' responses based on linguistic features and font size characteristics.

[1] https://www.nature.com/ncomms/

**Extract promotional language from academic papers.** To investigate the promotional intensity of authors regarding the innovative aspects of their work, we need to extract contribution-promoting sentences from the papers. Here, we define contribution-promoting sentences in academic papers as "sentences used to explicitly highlight the main contributions and innovative points of the research work." Although authors tend to emphasize the key points of their research in the title and abstract, they may still lack descriptions of innovative aspects in some detailed parts. In the introduction of a paper, authors clarify the research background while explicitly stating the specific problems to be solved or the core themes of the research, and they articulate the purpose and significance of the study, promoting the main innovative points of the research. Therefore, we selected the introduction of the paper as the corpus for extracting contribution-promoting sentences. Additionally, referencing the classification method of academic paper innovation by Leahey et al.(2023), we categorized the innovation in academic papers into theoretical innovation, methodological innovation, and result innovation, with specific definitions of each type of innovation as shown in Table 3.

**Table 3.** Definition of Three Innovation Types

| Innovation Type | Definition |
|---|---|
| Theoretical Innovation | Refers to breakthroughs in theoretical frameworks, models, or concepts. This can involve new theoretical perspectives, redefinitions of concepts, or extensions of existing theories, which advance the understanding and development of the discipline. |
| Methodological Innovation | Involves improvements or innovations in research methods, techniques, or tools. This can include new experimental designs, data collection methods, and analytical techniques, making research more |

[2] https://pymupdf.readthedocs.io/en/latest/

| | |
|---|---|
| | efficient and reliable, or enabling the resolution of previously unsolvable problems. |
| Result Innovation | Refers to new findings or conclusions obtained from the research. This type of innovation emphasizes the new knowledge or data gained from the research and its potential applications, which can have a significant impact on theory, practice, or policy. |

When extracting contribution-promoting sentences from academic papers, we employed a large language model. Specifically, we used DeepSeek-V3, a Mixture-of-Experts (MoE) language model with 671 billion total parameters, of which 37 billion parameters are activated for each token. DeepSeek-V3 has demonstrated strong performance across multiple benchmark datasets compared with other open-source models (DeepSeek-AI et al., 2024). We referenced the prompt templates provided by DeepSeek's official documentation to craft corresponding prompts that define innovative contribution sentences, instructing the large model to extract the original text of innovative contribution sentences from the titles, abstracts, and introductions of papers. To investigate the innovative points that academic paper authors focus on, we directed the large model to extract contribution-promoting sentences and categorize them into theoretical innovation, methodological innovation, and research outcome innovation, with each contribution-promoting sentence belonging to only one category. The dictionaries used to define the three innovation types are provided in Tables A4–A6 in the Appendix. After completing the prompt, we tested its extraction performance, randomly selecting 10 papers after each extraction test to observe the results, ensuring that all contribution-promoting sentences in the papers were extracted and assigned to the correct category, and that these sentences were sourced from the original text rather than generated by the model. Once the extraction performance met our expectations, we used the refined prompt to extract and classify contribution-promoting sentences from the introductions of papers across the entire dataset. The final prompt we used is shown in Table A2 in the Appendix.

**Extract innovative evaluation sentences from peer review**. To investigate reviewers' opinions on academic papers, we developed a "rule-based + machine learning" method to extract innovation evaluation sentences from review comments. Initially, we referenced the work of Leahey et al.(2023) on extracting innovation evaluation sentences from academic papers, using a large language model to generate common templates for innovation evaluation in peer-review comments. We also conducted a survey of language patterns in existing peer review corpora to develop and refine a lexicon for the preliminary extraction of innovation evaluation sentences, as detailed in Table A3 in the Appendix. However, due to the polysemous nature of innovation indicator words in peer-review comments—for example, 'original' can mean innovative or refer to the reviewer's initial opinion when placed before 'review'—we established corresponding rules. For instance, if 'new' is followed by words such as 'version', 'fig', or 'review', it is not recognized as a candidate for an innovation sentence. Based on this lexicon and these rules, we used regular expressions to extract candidate innovation sentences from peer-review comments, initially extracting 108,033 sentences containing innovation indicators from 15,328 peer-review comments.

To address the issue of low recall in the rule-based extraction method, we employed a machine learning approach to further classify the initially extracted innovation evaluation sentences. Specifically, we utilized SciBERT as our classification model to determine whether the preliminarily extracted innovation evaluation sentences were genuinely related to innovation. SciBERT is a pre-trained model based on the BERT architecture, optimized specifically for scientific texts and currently applicable to various natural language processing tasks such as text classification, named entity recognition, and question-answering systems, particularly in scientific applications (Beltagy et al., 2019). We randomly selected 1,500 candidate innovation evaluation sentences for manual annotation, distinguishing between innovation evaluation sentences and non-innovation evaluation sentences. The annotation was performed by two graduate students in library and information science, with a unified definition of innovation evaluation sentences in peer-review comments as "sentences in which reviewers make positive or negative evaluations about the innovation of the paper's theory, methods, results, etc." The Kappa coefficient calculated after annotation was 0.92. The dataset was divided into training, validation, and test sets with a ratio of 8:1:1, where the test set (10% of the data) is used to evaluate the model's performance on unseen samples. Ultimately, our extraction model achieved an accuracy of 0.88, a recall of 0.94, and an $F_1$ score of 0.92 on the test set.

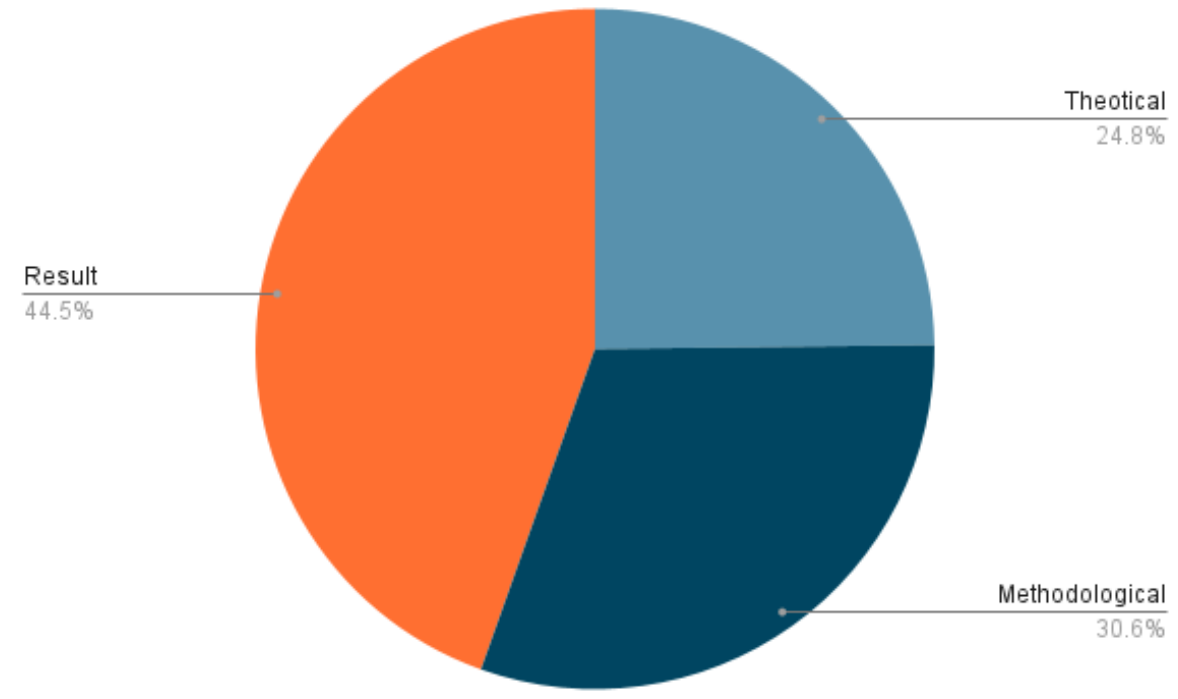

**Figure. 2.** The Proportion of Three Types of Innovation Evaluation Sentences

We employed SciBERT to classify the pre-extracted candidate sentences into two categories: innovation evaluation sentences and non-innovation evaluation sentences. Ultimately, we extracted 38,561 innovation evaluation sentences from all peer-review comments and categorized them into three types: theoretical innovation, methodological innovation, and result innovation. The classification rules are detailed in Tables A6, A7, and A8 in the Appendix, and the extraction results are illustrated in Figure 2. Not every innovation evaluation sentence falls into one of these three categories, as some sentences provide an overall assessment of the paper's innovativeness. For example, the sentence "*In my opinion, the novelty of this work is enough to guarantee a publication* *in Nature Communications.*" expresses a positive evaluation of the paper's overall innovativeness without specifying any particular aspect of innovation.

After completing the extraction and classification of peer-review comments, we calculated the innovation evaluation scores given by peer reviewers to the papers using the extracted innovation evaluations. Specifically, we conducted a survey of common linguistic forms of innovation evaluations in peer-review comments and constructed a lexicon of positive evaluations and a lexicon of negative innovation evaluations, as detailed in Tables A7 and A8 in the Appendix. The positive innovation lexicon includes sentiment words such as 'highly' and 'important', indicating that reviewers highly affirm the paper's innovativeness; whereas the negative innovation evaluation lexicon includes negative sentiment words such as 'lack' and 'insufficient', suggesting that reviewers find some aspect of the paper lacking in innovation. If an innovation evaluation sentence contains a positive innovation word, it is scored as 2 points; if it contains a negative innovation evaluation, it is scored as -1 point; all other ordinary innovation evaluation sentences are scored as 1 point. We summed the scores of each type of innovation evaluation sentence for each paper and calculated the average score across reviewers to obtain the innovation evaluation score for each paper.

### 3.2 Calculating the novelty of academic papers.

Novelty evaluation is vital for the promotion and management of innovation(Zhao & Zhang, 2025). This study employed the Novelty_U metric proposed by Uzzi et al., (2013) to measure the innovativeness of academic papers. This method is based on Schumpeter's (Chen et al., 2024) theory of combinatorial innovation, interpreting the innovation of a paper as a new combination of knowledge units and using the references of academic papers as a proxy for their knowledge sources to calculate the paper's novelty. The combination of knowledge, especially the combination of different types of knowledge, often produces novel knowledge(Chen et al., 2024).

Specifically, this novelty calculation method quantifies the degree of innovation of an article by using the atypicality of the journal combinations to which the references of the academic paper belong. Firstly, the journals to which the references in each paper belong are paired two by two, and the frequency of occurrence of each journal pair is counted. Then, the references from the same year are recombined, ensuring that the length and temporal distribution of the references for each paper remain unchanged. This step is repeated multiple times to obtain the frequency of occurrence of each reference pair under random conditions. Finally, the atypicality z-score is calculated based on the actual occurrence and the frequency of occurrence under random conditions for each journal pair.

$$\text{z-score} = (obs_{ij} - rand_{ij})/std_{ij} \quad (1)$$

$$\text{Novelty_U} = - P_{10}\ (\text{z-score}) \quad (2)$$

In this context, $obs_{ij}$ represents the actual frequency of occurrence of two journals in a paper, $rand_{ij}$ is the expected frequency of occurrence of the two journals in a paper, and $std_{ij}$ is the standard deviation of the occurrence of the two journals in a paper. Following the approach of Uzzi et al.(2013), we define the innovation of a paper as the negative value of the 10th percentile ($P_{10}$) of the z-scores of the reference journal pairs in the paper, sorted from smallest to largest. This means that the larger the value of Novelty_U, the more innovative the paper is considered to be.

### 3.3 Assessment of the intensity of promotional language.

To promote transparency in academic communication and ensure that research findings are presented in a truthful and reasonable manner, we evaluated the intensity of promotional language extracted from the introductions of academic papers. For the intensity of promotional language in the introduction, we referred to the work of Sun et al.(2024) and designed two metrics. The first metric is the proportion of promotional language in the introduction of the paper (PL), which represents the amount of effort the authors have spent on promotion in the introduction. The second metric is the proportion of promotional words in the introduction of the paper (PW). Promotional words refer to those with strong emotional connotations or exaggerated effects, typically used to attract attention, stimulate interest, or enhance the perceived value of something.

$$PL = \text{Num(words of promotional language)} / \text{Num(words of Introduction)} \times 100\% \quad (3)$$

$$PW = \text{Num(promotional words)} / \text{Num(words of Introduction)} \times 100\% \quad (4)$$

In this study, we focused on innovation-related promotional words such as 'new', 'unique', 'revolutionary', etc., which emphasize the innovative value of research findings. When constructing the promotional word lexicon, we referred to the lexicon proposed by Millar et al., (2022) based on promotional language in NIH grant applications, which includes 139 scientific promotional words. We further constrained the lexicon by treating words as promotional only when they appeared in the extracted promotional language, thereby reducing bias due to polysemy. Although the research corpus we used consists of the introduction sections of academic papers, the purpose of the promotional language is similar to that of grant applications, both aiming to promote their research to reviewers or readers. Therefore, based on Millar's lexicon, we manually surveyed the promotional language extracted using a large language model and added some commonly used promotional words in academic papers.

We used these two metrics as proxies for promotional intensity and, in conjunction with the reference-based innovation metric for academic papers, investigated what level of promotional intensity could garner more positive evaluations from reviewers at the same level of innovativeness. Combining the reviewers' innovation scores calculated in section 3.1, we compared the top 5% and bottom 5% of papers based on their Novelty_U scores to observe how the use of promotional language in papers with different levels of novelty affects innovation evaluations in peer reviews.

### 3.4 Reviewer Innovativeness Scoring and Disagreement Measurement.

To compare the degree of disagreement among multiple reviewers of the same paper regarding its innovativeness, we quantify divergence using the variance of reviewers' innovation scores. A larger variance indicates greater disagreement, whereas a smaller variance indicates greater consensus.

Our procedure proceeds in two steps. First, we compute an innovation score for each reviewer. We have already derived sentence-level innovation scores for individual review sentences via rule-based matching. A given reviewer may provide multiple sentences that comment on the paper's innovativeness, or none at all. If a reviewer offers no explicit assessment of innovativeness, we interpret this as tacit approval without sufficiently strong views to warrant a specific remark, and assign that reviewer an innovation score of 0; if a reviewer provides multiple sentences evaluating innovativeness, we take the mean of those sentence-level scores as the reviewer's innovation score. Formally, for reviewer *i*:

$$x_i = \begin{cases} 0 & n = 0 \\ \frac{\sum_{j=1}^{n} x_{ij}}{n} & n \neq 0 \end{cases} \tag{5}$$

Where, $x_i$ is reviewer i's innovativeness score, n is the number of innovation-related sentences provided by reviewer $i$, and $x_{ij}$ is the score for the j-th innovation-related sentence by reviewer i. After computing each reviewer's innovativeness score, we can calculate the variance of reviewers' assessments for manuscript X:

$$D(X) = \frac{1}{m}\sum_{i=1}^{m}(x_i - \bar{x})^2 \tag{6}$$

Where $D(X)$ represents the disagreement among different reviewers regarding the manuscript's innovativeness, m is the number of reviewers for manuscript X, and $\bar{x}$ is the average innovativeness score across reviewers.

### 3.5 Model Specification

To examine the relationship between the intensity of promotional language used in academic papers and peer review evaluations across different levels of novelty, in this study, we divide papers into three equal groups based on their novelty scores sorted from high to low: high-innovation group, medium-innovation group, and low-innovation group. Using PL and PW as proxies for promotional intensity, we conduct separate regression analyses for each group.

$$PRscore = \beta_0 + \beta_1 PL + \beta_2 PW + \beta_3 [Controls] + \varepsilon_i \tag{7}$$

Where i denotes individual papers, $\beta_0$ is the intercept, $\beta_1$ and $\beta_2$ are the coefficient capturing the marginal effect of promotional intensity (PL or PW) on the review score, $\beta_3$ is a vector of coefficients for the control variables, and $\varepsilon_i$ is the idiosyncratic error term.

We incorporate a series of control variables that may influence innovation evaluation outcomes into our analysis. These include: (1) the number of authors, to account for the potential effects of collaboration scale; (2) the length of the introduction, to control for the scope of textual exposition; (3) the publication year, to capture temporal trends and cohort effects; and (4) the disciplinary field, to adjust for domain-specific norms and evaluation practices.

# 4 Result

In this section, we present the differences in innovation focus between paper authors and reviewers, and analyze the impact of promotional language in academic papers on the peer review process in conjunction with innovation metrics.

## 4.1 Innovation Focus

Here, we utilize the contribution description sentences extracted from academic papers and the innovation evaluations extracted from peer-review comments to investigate *RQ1*, which is whether paper authors and reviewers place more emphasis on theoretical innovation, methodological innovation, or result innovation in scientific research within the field.

**Innovation Focus of author.** In this section, we conduct a statistical analysis to determine which aspects of innovation paper authors are more focused on. We perform a statistical analysis on the various types of contribution-promoting sentences extracted using the large model. The statistical results are shown in Table 4, with the proportion of papers containing result innovation being the highest at 87.19%. This indicates that paper authors place greater emphasis on the innovation of results and are more likely to directly promote the innovation of their research findings in the introduction section of their papers.

**Table 4.** Total Number and Proportion of Three Promotion Types

| Promotion Type | Number | Proportion |
|---|---|---|
| Theoretical Innovation | 5817 | 37.80% |
| Methodological Innovation | 4619 | 30.01% |
| Result Innovation | 13418 | 87.19% |

From the proportion of various types of innovation, we can see that in some papers, authors claim that their research encompasses multiple types of innovation. We have counted the number of such papers and found that 33.4% of the papers contain at least two or more types of innovation-related contribution promotions. Among these, the proportion of papers that include all three types of innovation is the highest, at 26.46%. This allows us to explain why result innovation accounts for the largest proportion, namely that when paper authors make theoretical or methodological innovations, they often accompany these with innovative results. Authors of these papers believe that when declaring their contributions, they are comprehensively promoting the contributions of their research in all aspects. However, the majority of authors only promote the innovative results of their research in the introduction section, considering result innovation to be the core value of their study.

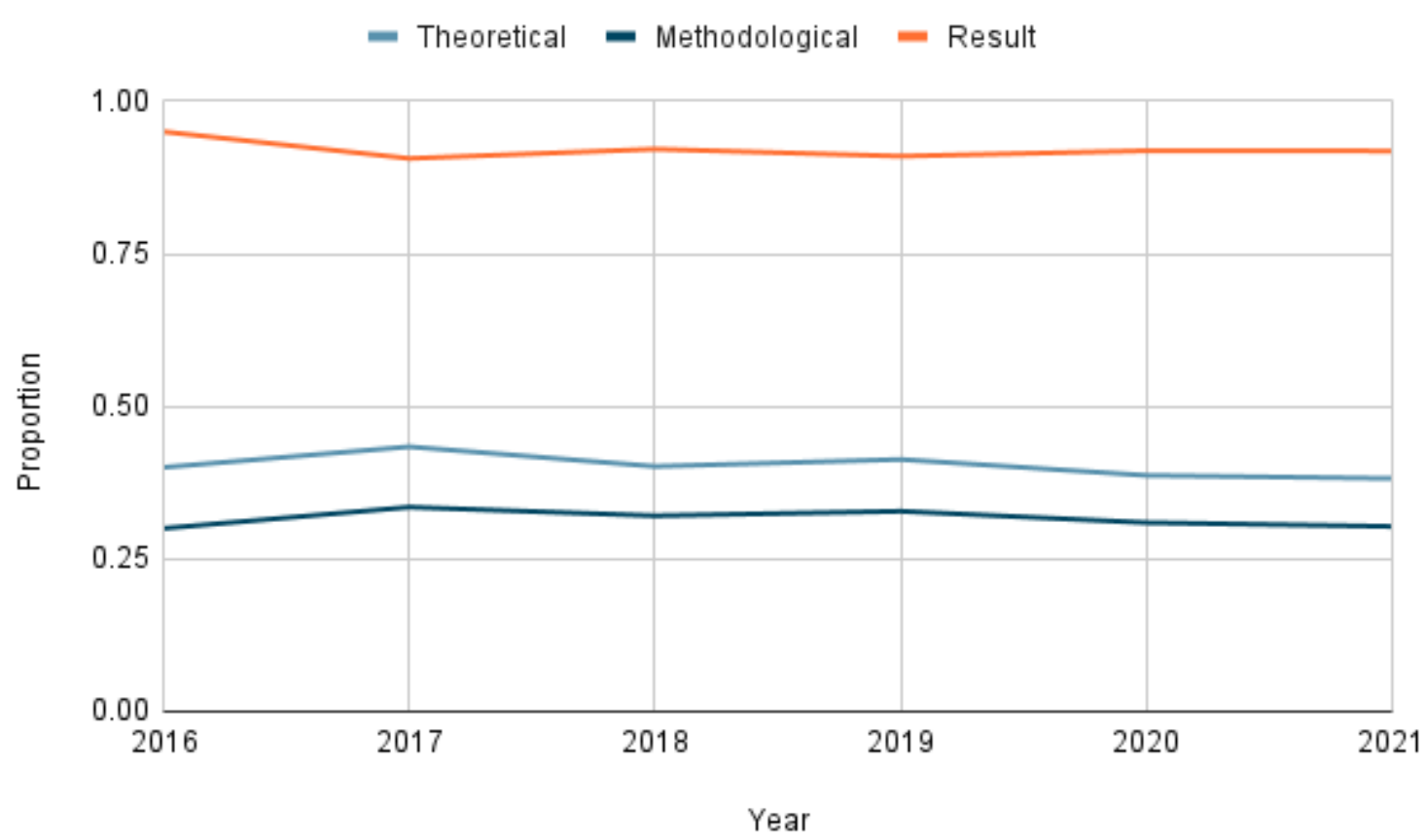

**Figure. 3.** Changing Innovation Focus of Authors Over Time

We have examined the trend in the emphasis paper authors place on promoting various aspects of innovation by incorporating the publication dates of the papers. As shown in Figure 3, we can observe that the proportion of these three types of contribution promotions has remained relatively stable in recent academic writing. Most authors declare the innovative results of their research in the introduction section.

Next, we investigated which aspects of innovation are more emphasized by authors in different fields based on the domain of the papers. The results, as shown in Figure 4, reveal that authors across all five fields in Nature Communications place greater emphasis on the innovation of results, and their focus on theoretical innovation is also slightly higher than that on methodological innovation.

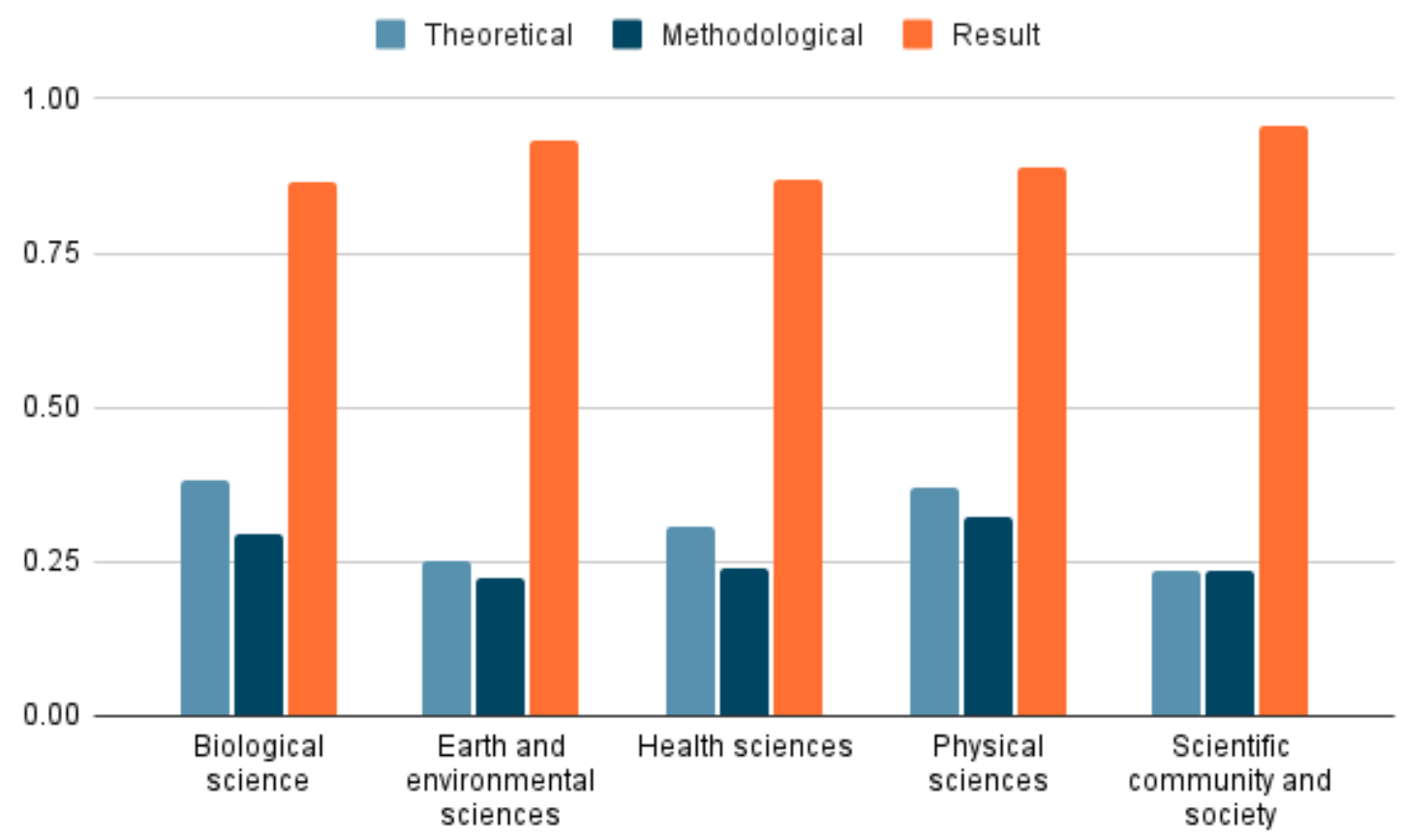

**Figure. 4.** Innovation Focus of Authors in Five Different Fields.

After extracting and statistically analyzing the contribution-promoting sentences in academic papers and the innovation evaluation sentences in peer-review comments, we found that both reviewers and authors of academic papers are more inclined to evaluate and promote the innovative results of the papers. Analyzing the publication dates of the papers, we observed that the focus on innovation in the writing and review process of academic papers has remained relatively stable in recent years. When dividing the papers into different disciplinary fields for study, we discovered that authors from various disciplines have similar perspectives on the angles of contribution promotion, while reviewers' focuses differ significantly. For example, in the fields of scientific community and society and physical science, reviewers pay more attention to methodological innovation than in other fields.

**Innovation Focus of reviewer.** Reviewers' evaluations of paper novelty are concentrated on research outcomes. As shown in Figure 2, among all extracted innovation evaluation sentences, those commenting on result innovation constitute the largest proportion (50%), followed by methodological (27.6%) and theoretical (22.4%) innovations.

Analyzing the overlap in the evaluation of review comments, we found that 36.21% of the peer-review comments for papers contained two or more types of innovation evaluations. Among these, 1,611 papers had all three types of innovation mentioned—theoretical, methodological, and result; 1,573 papers had both theoretical and result innovations mentioned; another 1,573 papers had both methodological and result innovations mentioned; and 686 papers had both theoretical and result innovations mentioned. This shows that the innovations in a paper often do not exist in isolation; for example, a paper may develop a new method and use it to discover new results. Reviewers tend to consider all aspects comprehensively when evaluating the innovation of a paper, which also explains why result innovation accounts for the largest proportion in innovation evaluations. This is because result innovation is more closely related to theoretical and methodological innovations, and when reviewers identify theoretical or methodological innovations, they are inclined to simultaneously evaluate the innovativeness of the results.

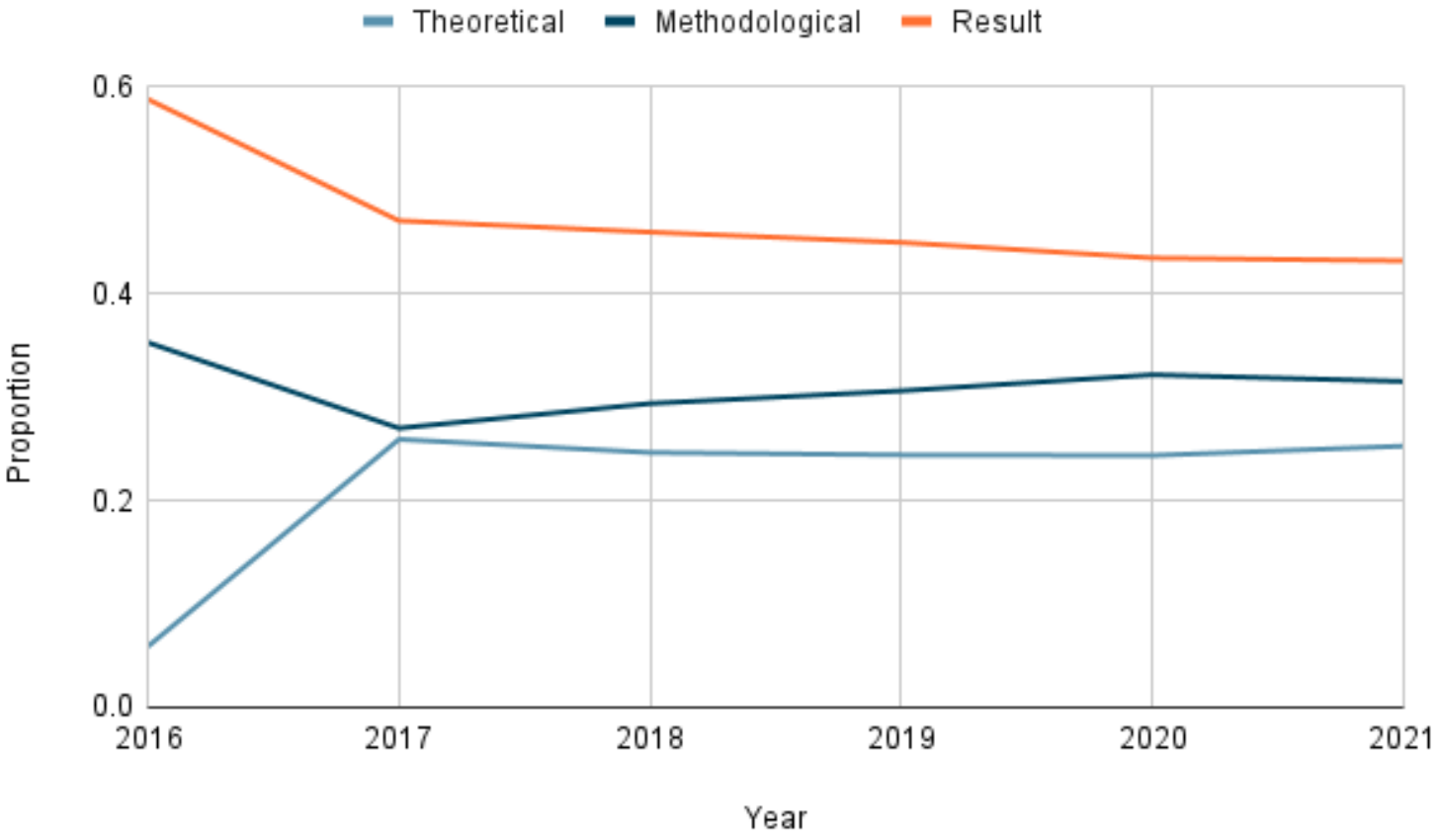

**Figure. 5.** The changing innovation focus of reviewers over time.

We examined the proportion changes of different types of innovation evaluation sentences over time by incorporating the publication dates of the papers. Figure 5 illustrates the proportion changes of various types of innovation evaluation sentences from 2016 to 2021. The noticeable change around 2017 is partly due to the fact that *Nature Communications* only began publishing peer-review reports in October 2016, resulting in relatively few observations for 2016 in our dataset. Overall, the proportions of the three types of innovation evaluation sentences have remained relatively stable, with the proportion of methodological innovation evaluations showing an upward trend, while the proportion of result innovation evaluations has been declining. This indicates that in recent years, reviewers have been placing increasing emphasis on the innovation of research methods.

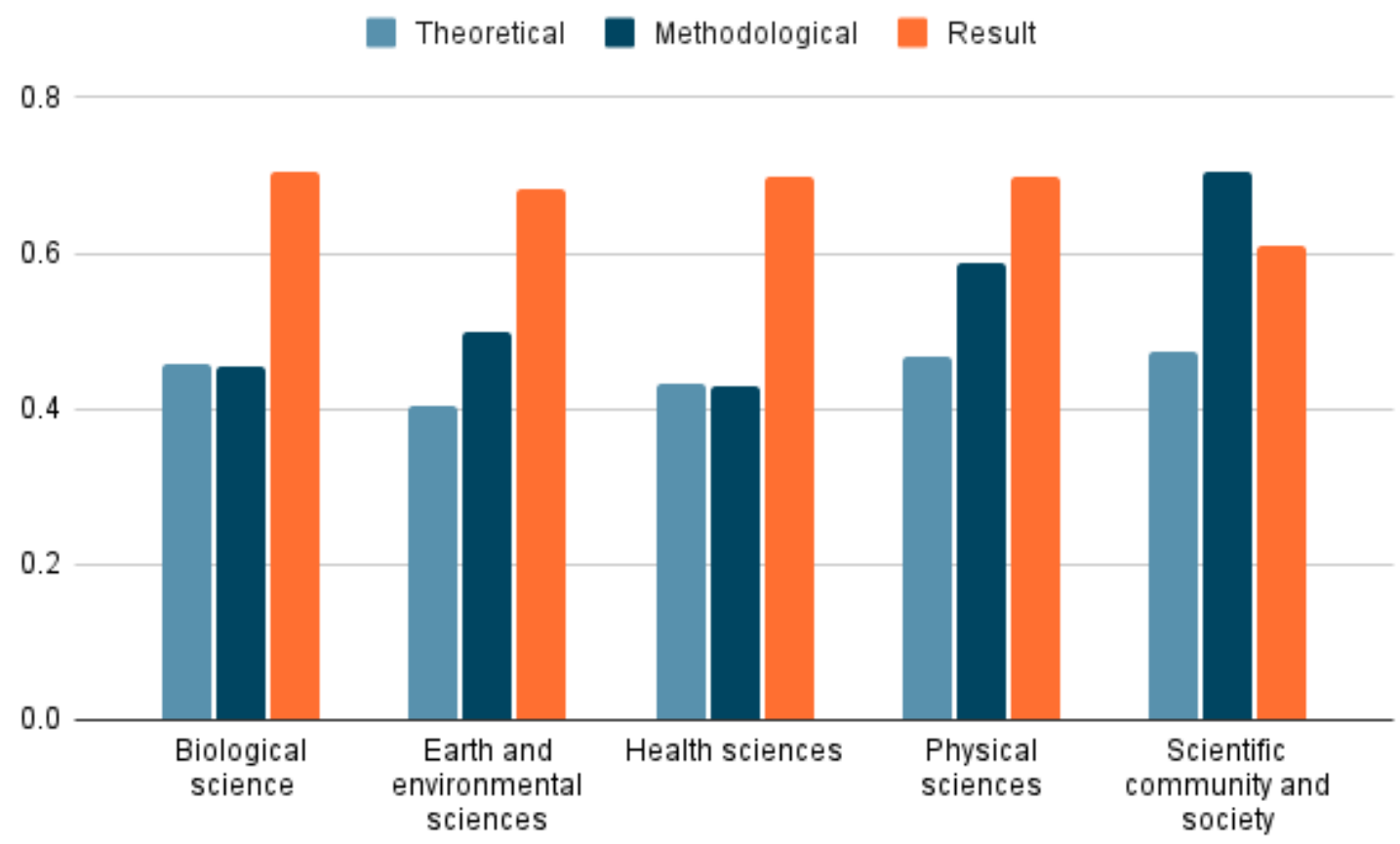

**Figure. 6.** Innovation Focus of Reviewers in Five Different Fields.

Next, we divided the study into five different disciplinary fields to examine the focus of reviewers on innovation in peer-review comments. The results, as shown in Figure 6, indicate that all disciplinary fields comprehensively review various aspects of innovation in papers, especially the innovation of results. In the field of scientific community and society, reviewers' attention to methodological innovation is particularly prominent, with 70.52% of the innovation evaluation sentences in the peer-review comments for papers in this field being assessments of methodological innovation. This is because this field is interdisciplinary, requiring timely follow-up and integration of new methods from various disciplines to address current social issues; whereas the other four fields mostly rely more on the specialized knowledge and techniques of their respective fields to drive the production of more innovative research outcomes.

### 4.2 Differences in Innovation Focus Between Authors and Reviewers

In section 4.1, we analyzed which aspects of innovation are focused on in academic papers and peer-review comments, respectively. It can be observed that although both paper authors and reviewers pay considerable attention to the innovation of paper results, their focuses still differ. Therefore, in this section, we address *RQ2*, which is what differences exist between authors and reviewers when evaluating the innovative points of a paper, and which innovations mentioned by authors in the introduction are recognized by reviewers?

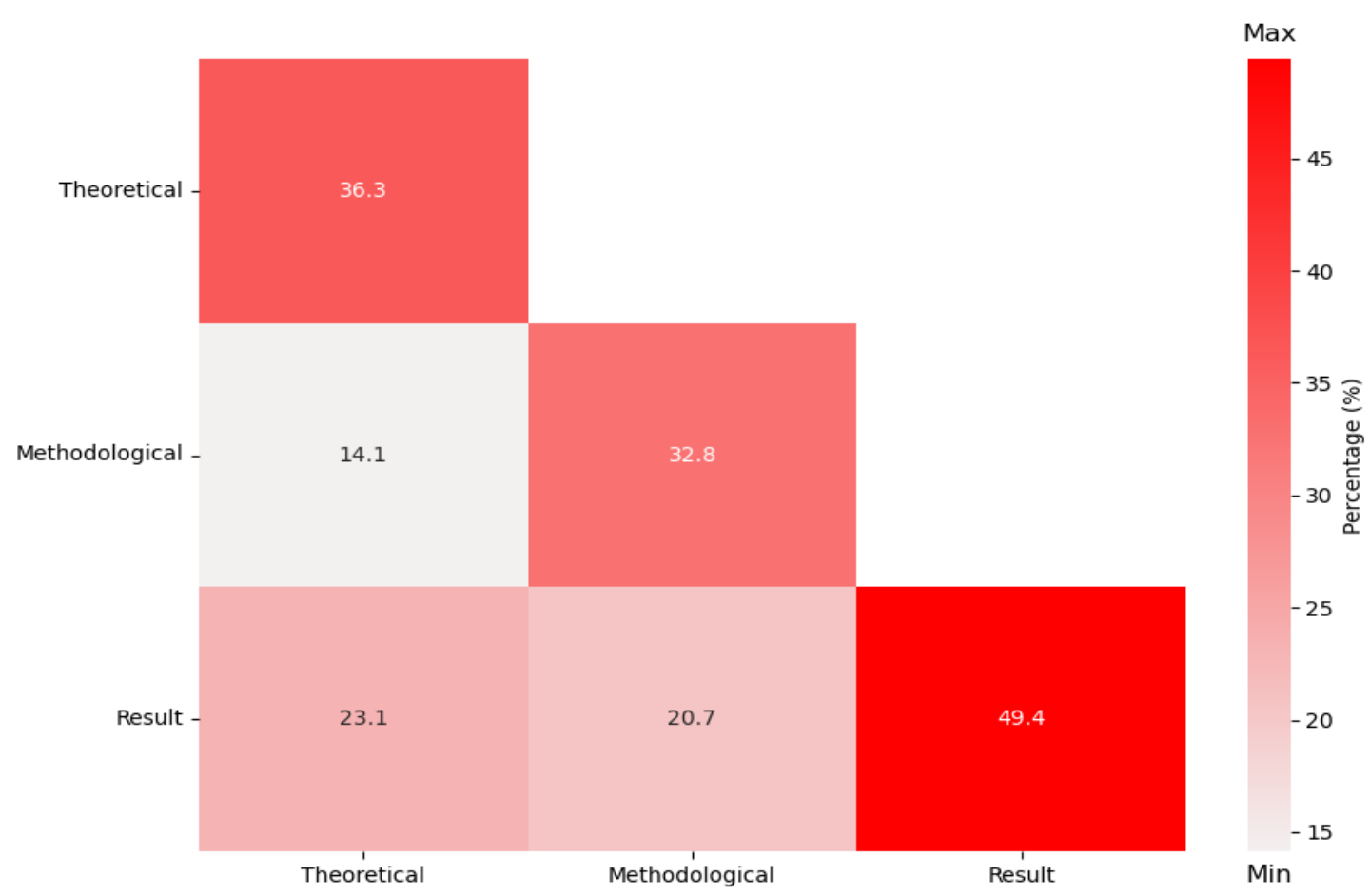

**Figure. 7.** Heatmap of Overlap Ratio Between Reviewers' Positive Evaluations and Authors' Promotional Language.

We investigated the relationship between the contribution-promoting sentences provided by authors and the innovation evaluations given by reviewers in the same academic paper, and used this to create a heat map. As shown in Figure 7, the horizontal axis represents the types of innovation described by the paper authors, and the vertical axis represents the types of positive innovation evaluations given by peer review experts. The content of each cell indicates the proportion of papers that received corresponding positive evaluations from review experts when authors provided that type of contribution-promoting sentence. For example, 49.4% of the papers that promoted their research result innovations were recognized by at least one positive evaluation on result innovation peer review experts; 20.7% of the papers that promoted their research methodological innovations also received positive evaluations for theoretical innovation from the review experts. From the figure, we can see that when describing research result innovations in a paper, it is more likely to receive positive evaluations from review experts compared to the other two types of innovation. Considering that a paper may contain multiple innovations and that review experts may also make innovation evaluations on various aspects of the paper, we can observe that theoretical innovation is closely linked with result innovation, with 23.1% of the papers proposing theoretical innovation and receiving positive evaluations from reviewers for their result innovations.

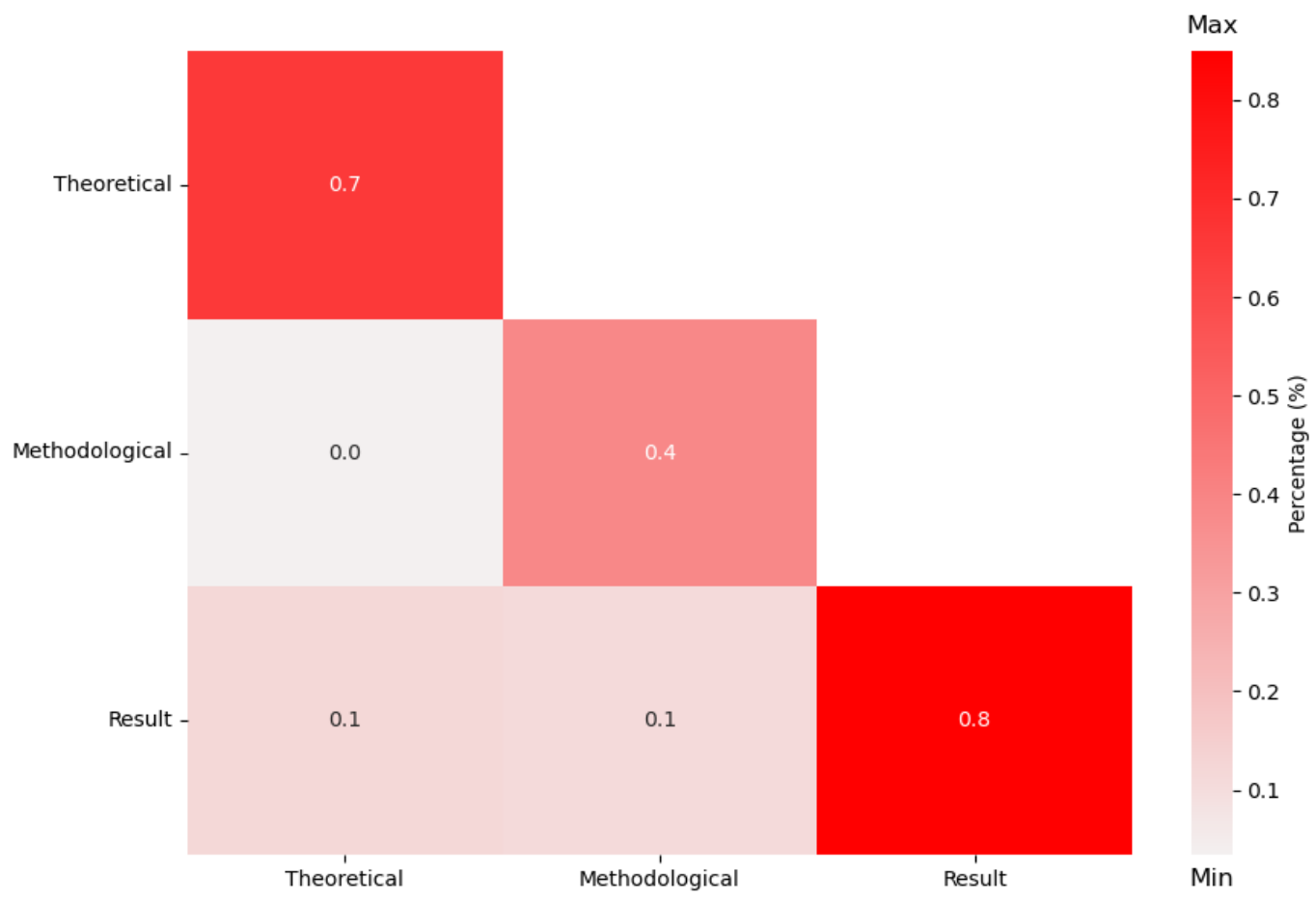

**Figure. 8.** Heatmap of Overlap Ratio Between Reviewers' Negative Evaluations and Authors' Promotional Language

However, during the review process, reviewers may also provide negative evaluations, such as considering that the paper is not as novel as claimed, or that similar topics have been studied before but the authors did not mention them. We have also conducted statistics on these papers, and the results are shown in Figure 8. Since the data we used only includes the original texts of accepted papers and peer review corpora, the quality is generally high, and there are fewer negative evaluations in the peer-review comments. Among them, negative evaluations mostly appear in the innovation points claimed by the authors themselves. For example, 0.8% of the authors promoted the innovativeness of their results, but the reviewers considered their results not to be innovative.

Through the above research, we can find that the innovation promotion in the introduction of a paper does indeed draw the attention of reviewers to that type of innovation. However, if the contribution is misrepresented or improperly promoted, it is more likely to be refuted by review experts. As for innovations not declared in the introduction, review experts rarely give negative opinions.

### 4.3 Inherent Novelty as a Driver of Promotional Language

We know that the promotional language in academic papers is used to explain the innovations of the research to readers. But to what extent is the promotional language in papers influenced by their inherent novelty? In this section, we investigate RQ3, examining whether more novel papers tend to use more promotional language and promotional words. We use Novelty_U to represent the novelty of the paper, and measure the promotional intensity by the proportion of promotional language in the introduction and the proportion of promotional words in the introduction.

**Table 5.** Correlation Between PL and Novelty_U

| Variable | Theoretical | Methodological | Result | Total | Novelty_U |
|---|---|---|---|---|---|
| Theoretical | 1.000 | .447*** | -.088*** | .804*** | .044*** |
| Methodological | .447*** | 1.000 | -.032*** | .624*** | 0.005 |
| Result | -.088*** | -.032*** | 1.000 | .445*** | .036*** |
| Total | .804*** | .624*** | .445*** | 1.000 | .052*** |
| Novelty_U | .044*** | 0.005 | .036*** | .052*** | 1.000 |

**Note**: *: $p < 0.05$, **: $p < 0.01$, ***: $p < 0.001$.

Firstly, we calculate the correlation between the proportion of various types of innovation-promotional language in the introduction and the novelty measured based on references. The results are shown in Table 5. It can be seen that novelty is positively correlated with the proportion of various types of innovation-promotional language in academic papers, and it is significantly correlated with the proportion of theoretical novelty claims and result novelty claims. This indicates that novel papers tend to use more space in the introduction to describe the innovations of their work.

**Table 6.** Correlation Between PW and Novelty_U

| Variable | Theoretical | Methodological | Result | Total | Novelty_U |
|---|---|---|---|---|---|
| Theoretical | 1.000 | .273*** | .026*** | .742*** | .037*** |
| Methodological | .273*** | 1.000 | 0.012 | .480*** | -0.012 |
| Result | .026*** | 0.012 | 1.000 | .626*** | 0.005 |
| Total | .742*** | .480*** | .626*** | 1.000 | .023** |
| Novelty_U | .037*** | -0.012 | 0.005 | .023** | 1.000 |

**Note**: *: $p < 0.05$, **: $p < 0.01$, ***: $p < 0.001$.

Next, we calculate the correlation between the proportion of promotional words in the introduction and novelty. Only promotional words within the promotional language are included in our statistics. That is, a higher proportion of promotional words indicates a greater promotional effort by the authors and more confidence in the novelty of their research. According to the correlation results in Table 6, novelty is positively correlated with the proportion of theoretical promotional words and result promotional words, but negatively correlated with the proportion of methodological promotional words. This shows that innovative papers tend to use more strongly promotional words that emphasize the novelty of their research to describe their theoretical innovations in the introduction, while using fewer such words to describe their methodological innovations.

A possible explanation is that for highly novel papers, methodological advances might be presented as a means to achieve the novel results or as self-evident, thus requiring less emphatic promotion compared to the theoretical underpinnings or the results themselves.

### 4.4 The Relationship Between Promotional Intensity and Peer Review

Having established the correlation between a paper's inherent novelty and peer review outcomes, we now address RQ4: How does the promotional intensity employed by authors influence peer review evaluations? This allows us to explore more deeply the impact of authors' writing strategies on the review process. reviewers' judgments of a paper's innovativeness are significantly correlated with the paper's actual novelty. Therefore, we divided the papers into three equal groups based on their innovativeness scores (high, medium, and low) and performed subgroup regression analyses. The results are presented in the table below.

**Table 7.** Regression Results of Promotional Intensity on Reviewer Novelty Scores

| **Variables** | **Dependent Variable: PRscore** | | | | | |
|---|---|---|---|---|---|---|
| | **(1)** | **(2)** | **(3)** | **(4)** | **(5)** | **(6)** |
| Novelty | Low | Medium | High | Low | Medium | High |
| PL | 0.0058 | 0.0007 | 0.0342*** | | | |
| PW | | | | 0.0126 | 0.0000 | 0.0268** |
| Author count | Yes | Yes | Yes | Yes | Yes | Yes |
| Field | Yes | Yes | Yes | Yes | Yes | Yes |
| Publish year | Yes | Yes | Yes | Yes | Yes | Yes |
| Observations | 3704 | 3703 | 3704 | 3704 | 3703 | 3704 |
| $R^2$ | 0.006 | 0.008 | 0.0129 | 0.0067 | 0.0080 | 0.0124 |

**Note**: **: $p < 0.01$, ***: $p < 0.001$.

The regression results are shown in Table 7. The analysis reveals that only in the high-innovativeness group do both promotional language proportion (PL) and promotional word proportion (PW) exert a significant positive effect on reviewer evaluations. In contrast, for papers with low or medium innovativeness, promotional intensity shows no statistically significant impact. In other words, if a paper's inherent novelty is limited, employing more promotional language does not translate into more favorable reviewer assessments. Conversely, for highly innovative papers, stronger promotional language in the introduction is associated with more positive evaluations from reviewers.

A plausible explanation for this pattern is that when a paper's novelty is inherently low, reviewers can more readily identify its shortcomings and provide a more objective assessment of its innovativeness. For highly novel papers, however, authors' promotional language serves to guide reviewers toward understanding the paper's innovative contributions, thereby eliciting higher novelty ratings.

### 4.5 Promotional Language and Reviewer Disagreement in the Gray Zone

To gain a deeper understanding of the relationship between promotional language in academic papers and the peer review process, we address RQ5 here: whether promotional language in papers is associated with discrepancies among reviewers regarding the perceived novelty of the work during the review process. To investigate this, after extracting sentences related to novelty evaluation from the review comments, we calculate the level of disagreement among reviewers concerning their assessments of the paper's novelty. From RQ3 and RQ4, we know that the promotional language used in papers is related to their inherent novelty, and depending on the level of innovation of the paper itself, there may be varying correlations between the authors' promotional language and the innovation-related feedback from peer reviewers. Therefore, we sorted the paper data in descending order based on their inherent novelty (Novelty_U), calculated the correlation between the promotional language and disagreement between reviewers every 5%, and visualized the results.

As shown in Figure 9, the X-axis represents the percentile of papers sorted by Novelty_U in descending order, while the Y-axis represents the correlation coefficient between promotional intensity (PW) and disagreement between reviewers. The red stars indicate the significance levels of the correlations. We found that for both highly innovative and minimally innovative papers, there is a weak correlation between promotional intensity and disagreement between reviewers; however, for papers with moderate innovation levels, there is a significant correlation between promotional intensity and disagreement between reviewers. In other words, for moderately innovative papers, greater promotional intensity tends to lead to greater disagreement between reviewers.

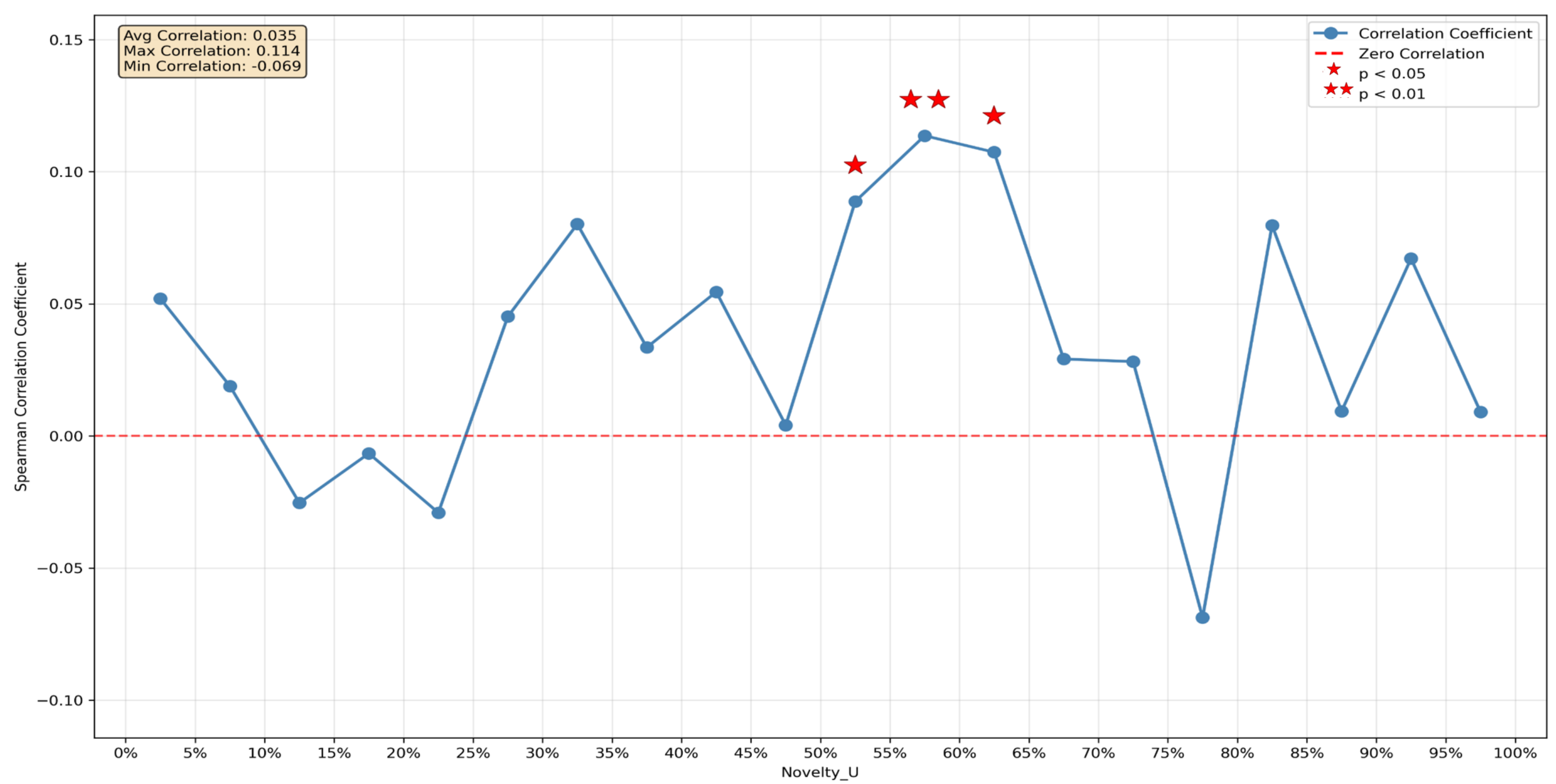

**Figure. 9.** Trend of Correlation between Promotional Intensity and Disagreement between Reviewers with Novelty_U

A possible explanation for this phenomenon is that for papers with either extremely high or extremely low novelty, the inherent evidence is sufficiently strong or weak to the extent that promotional language can hardly sway reviewers' judgments based on factual evidence. Conversely, for papers with moderate innovativeness, the evidence itself contains ambiguity and allows room for interpretation. In such cases, promotional language serves as a critical "external cue" that influences reviewers' assessments.

# 5 Discussion

## 5.1 Implications

We will elaborate on the implications of this study from both theoretical and practical perspectives.

**Theoretical implications**. This study explores the cognitive gap between authors and peer reviewers regarding the perception of novelty in academic papers, specifically using *Nature Communications* as a case study. Despite the availability of open peer review datasets, these resources often lack standardized writing conventions, which complicates the extraction of meaningful insights.

First, we developed and validated an analytical framework for examining the cognitive gap between authors and reviewers regarding the novelty of academic papers. Unlike previous studies that often focus on a single perspective (e.g., analyzing only author promotion or only review comments), this study systematically integrates authors' promotional language in the introduction and reviewers' novelty assessments in the peer review process within a unified analytical framework for the first time. By categorizing innovations into theoretical, methodological, and result dimensions, we conducted a fine-grained analysis of cognitive differences, revealing the misalignment between authors and reviewers. This provides new insights into cognitive transmission and evaluative feedback in the scholarly publishing process.

Second, we offered an in-depth explanation of the effectiveness of promotional language. Rather than treating its effect as linear or universal, this study tested its boundary conditions by incorporating the inherent novelty of papers into the analysis. The findings indicate that the impact of promotional language depends on the paper's own novelty level, exerting a significantly positive influence on review outcomes only for highly novel papers. This suggests that promotion is not merely a rhetorical device independent of research quality, but rather serves as a bridge linking the paper's intrinsic novelty with reviewers' perceptions.

Third, we uncovered the fragility of consensus in the "middle ground," providing a new perspective for understanding reviewer disagreement. Innovatively shifting the unit of analysis from individual reviewer scores to disagreement among reviewers, we found that the amplifying effect of promotional language on disagreement is concentrated in the "gray zone" of moderately novel papers. This reveals that when a paper's novelty is inherently ambiguous, promotional language acts as a critical external cue, exacerbating differences in reviewers' interpretations of innovativeness. This not only extends research on promotional language to its impact on the stability of the review process, but also furnishes evidence-based theoretical explanation for the emergence of controversy in peer review—suggesting that disagreement may not stem from review errors, but rather from cognitive divergence among experts in situations of evidential ambiguity.

**Practical implications**: The experimental conclusions of this study can provide recommendations for researchers in academic paper writing. To better enable review experts to understand the innovative aspects of the paper, authors should articulate the corresponding points of innovation in the introduction section, but also avoid over-promotion. This is because while promotional language in the introduction can benefit genuinely innovative parts, improper promotion may also raise doubts among review experts.

After completing the writing of their papers, authors can refer to existing academic paper innovation metrics to estimate the novelty of their own research. If the research is highly novel, more promotional language and words can be used in the introduction to make reviewers more aware of the innovative aspects of the research; if the novelty is low, the use of aggressive promotional words should be avoided to prevent exaggeration from raising doubts among reviewers.

Our work also provides a valuable lens for editors and the broader scholarly community. Editors can consider promotional intensity as a factor when interpreting conflicting reviews. Furthermore, by highlighting this dynamic, we hope to encourage a culture where the substance of evidence is prioritized, leading to a more efficient and transparent review system for everyone.

### 5.2 Limitations

This study has certain limitations. Firstly, the effectiveness of extracting contribution-promoting sentences from the original academic papers and innovation evaluation sentences from peer-review comments needs improvement. The accuracy of the extraction results may affect the validity of subsequent conclusions. Particularly, the lack of uniform standards in peer-review comments, the varying language styles of reviewers, and some implicit evaluations of paper innovation have impacted our extraction accuracy. Secondly, the corpus we used is limited to papers published in *Nature Communications*. Although this journal covers multiple disciplines within the natural sciences, the findings of this study have not been fully validated in some disciplinary fields. Moreover, different journals may have different review requirements, and the focus of reviewers on innovation may change with alterations in review criteria. However, currently, only a minority of scientific publications choose to open peer-review comments, so this study has only made a preliminary exploration of the research question using papers published in *Nature Communications*. Lastly, the dataset used in this study consists solely of accepted papers and their peer-review comments, and does not include data from rejected papers, which may also affect the generalizability of the study's results.

## 6 Conclusion and future works

In this paper, we investigated the cognitive differences between paper authors and reviewers regarding the innovation of papers during the writing and review process, provided recommendations for academic paper writing, promoted appropriate promotion in the academic paper writing process, and reduced the cognitive gap between paper authors and review experts. While both authors and reviewers prioritize result innovation, reviewers adopt a more comprehensive evaluative lens, paying significant attention to theoretical and methodological advancement alongside results. Authors, in contrast, direct much of their promotional emphasis toward their reported findings. Furthermore, the effectiveness of promotional language is not universal but is critically depend on the paper's inherent novelty. For highly novel papers, stronger promotional intensity correlates with more positive reviewer evaluations. However, for papers with lower inherent novelty, excessive promotion backfires, leading to lower novelty ratings. Most notably, promotional language exerts its most potent influence in the "gray zone" of moderate innovativeness, where it significantly amplifies disagreement among reviewers.

Future work will focus on improving the extraction accuracy for contribution and innovation sentences using advanced large language models. Since this study focuses on papers published in *Nature Communications*, the dataset mainly includes articles that have already passed the journal's novelty standards. Future studies could extend the analysis to other open peer-review journals, such as eLife or F1000Research, which may provide more diverse evaluation contexts. The dataset will be expanded to include rejected manuscripts to investigate if their outcomes are linked to promotional misalignment. Additionally, we will examine how these dynamics vary across different disciplinary fields.

## Acknowledgments

This paper was supported by the National Natural Science Foundation of China (Grant No.72074113) and Open Foundation of the ISTIC-CLARIVATE Analytics Joint Laboratory for Scientometrics (No. IC2502). This paper is an extended version of the ISSI 2025 paper "*Yang C., & Zhang, C. (2025). Examining the Cognitive Gap Between Authors and Peer Reviewers on Academic Paper Novelty. In: Proceedings of the 20th International Conference on Scientometrics and Informetrics (ISSI 2025), Yerevan, Armenia, 2025.*"

## Declarations

Conflict of interest: The author(s) declared no potential conflicts of interest with respect to the research, authorship, and/or publication of this article.

## Appendix: Dictionaries and Rules for Coding

**Table A1.** Example of three promotional types

| Promotion Type | Author description | Reviewer Comments |
|---|---|---|
| Exaggeration | This groundbreaking study introduces a revolutionary method that will completely transform data privacy in machine learning. | The claim that this method will 'revolutionize data privacy in machine learning' appears overly ambitious. For instance, Smith et al. (2021) demonstrated that while advancements in data privacy are significant, they often come with trade-offs in model performance and complexity. I recommend that the authors provide a more balanced view that acknowledges these trade-offs and the context in which their method may be effective. |
| Insufficient promotion | This study presents a approach to enhance data privacy protection in machine learning. While the method has demonstrated some effectiveness on certain datasets, it has not yet undergone extensive empirical validation. | To avoid understatement, more information about the advantages of the research method, specific experimental results, and its potential impact should be included in the description. |
| Appropriate promotion | This study presents a novel approach that demonstrates notable improvements in data privacy within machine learning frameworks compared to existing methods. Our results indicate enhanced protection of sensitive information while maintaining model performance. | The authors successfully provide a balanced description of their contributions, clearly articulating the improvements over existing methods without relying on hyperbolic claims. |

**Note**: The parts marked in red indicate promotional language

**Table A2.** Prompt Used in Promotional Language Extraction

| | |
|---|---|
| | **The persona pattern:**<br>You are a proficient linguist skilled in reading academic articles.<br>**Introduce the target of our task:**<br>You will be given a paragraph in the Introduction section of a publication. Please follow the instructions to label the paragraph in the Introduction section provided by user. In the introduction of a paper, the author mentions the innovations of the entire work, which we define as contribution statements.<br>**Definition of contribution statement:** |
| Instruction | Contribution statements are sentences or paragraphs in academic papers that clearly highlight the main contributions and innovations of the research work.<br>**Definition of three types of innovation:**<br>These contributions can be categorized into three types:<br>**(See Table 3.)**<br>**Introduce the details in our task:**<br>Each paper's introduction may contain one or more of these three types of innovations, and each sentence belongs to only one type of innovation. Please read the provided research paper's introduction carefully and extract the original sentences representing these contribution statements, categorizing them into theoretical innovation, methodological innovation, and result innovation. No explanations are needed for the extracted results. |
| Input | Introduction of an article. |
| Output | **theoretical innovation**: [extracted theoretical innovation statement](if none, leave blank);<br>**methodological innovation**: [extracted methodological innovation statement](if none, leave blank);<br>**result innovation**: [extracted result innovation statement](if none, leave blank)" |

**Table A3.** Innovation Signifying Terms Dictionary

| New | | | |
|---|---|---|---|
| novel | new | innovative | creative |
| novelty | uncover | fill the knowledge gap | methodological step |
| breakthrough | groundbreaking | pioneering | trailblazing |
| disruptive | revolutionary | unprecedented | advancement |
| introduce | propose | unique | originaL |

When these words appear after the word “new” or ‘original’, the word “new” or ‘original’ is not tagged as NEW: 'figure', 'we', 'our', 'table', 'version', 'fig', 'review', 'paragraph', 'claim', 'manuscript', 'comment', ' added', 'new text', 'sample', 'avoid', 'supp'.

We exclude mentions of “first” as NEW, unless one of these words appears immediately afterward: principle', 'result', 'observation', 'attempt', 'experiment', 'synthesis', 'study', 'comprehensive', 'application', 'description', 'describe', 'evidence', 'design', 'time'.

**Table A4.** Theoretical Innovation Terms Dictionary

| **Theoretical** | | | |
|---|---|---|---|
| concept | generalize | mechanism | synthesize |
| theoretical | explanation | hypothesis | model |
| term | theory | insight | point |
| idea | hypotheses | thesis | explain |

**Table A5.** Methodological Innovation Terms Dictionary

| **Methodological** | | | |
|---|---|---|---|
| method | analysis | classification | experiment |
| methodology | strategy | analyze | criteria |
| formula | procedure | technique | apparatus |
| criterion | index | process | technology |
| design | means | protocol | tool |
| equipment | measure | quantify | calculate |
| test | examine | experimental | approach |

**Table A6.** Result Innovation Terms Dictionary

| Result | | | |
|---|---|---|---|
| structure | confirm | finding | observation |
| response | correlation | found | outcome |
| result | effect | discovery | prove |
| support | evidence | identify | rate |
| show | observation | data | report |

If a sentence contains two or more different terms from the innovation aspect dictionaries, we conduct a syntactic analysis to observe which terms from Table A3 modify the terms from Table A4 to A6, thereby determining the innovation type of the sentence.

**Table A7. Positive Innovation Terms Dictionary**

| Positive | | | |
|---|---|---|---|
| highly | interest | completely | important |
| strong | indeed | surprise | extensive |
| sound | timely | extremly | incredibly |
| remarkably | exceptionally | significantly | thorough |
| thoughtful | clever | unquestionably | robust |

**Table A8. Negative Innovation Terms Dictionary**

| Negative | | | |
|---|---|---|---|
| lack | insufficient | inaccurate | ineffective |
| unconvincing | inappeoriate | slight | |

When the word “not, “no,” or “lack” appears near the word “new,” it is tagged as negative.